\documentclass[aps,prd,twocolumn,nofootinbib,
showpacs,preprintnumbers,amsmath,amssymb]{revtex4}
\usepackage[dvips]{graphicx}
\usepackage{enumerate}
\usepackage{longtable}

\newcommand{\beq}{\begin{eqnarray}}
\newcommand{\eeq}{\end{eqnarray}}
\newcommand{\non}{\nonumber\\}
\newcommand{\SU}{\text{SU}}
\newcommand{\SO}{\text{SO}}
\newcommand{\U}{\text{U}}

\def\6#1{{\underline{#1}}}
\def\m6#1{{\underline{#1}\,}}

\newdimen\Tdim
\def\ispan{{\setbox0=\hbox{i}%
\Tdim\ht0\advance\Tdim\dp0\rule[-\dp0]{0pt}{\Tdim}}}
\def\jspan{{\setbox0=\hbox{j}%
\Tdim\ht0\advance\Tdim\dp0\rule[-\dp0]{0pt}{\Tdim}}}
\def\Tspan#1{{\setbox0=\hbox{#1}%
\Tdim\ht0\advance\Tdim\dp0\advance\Tdim.55ex\rule[-\dp0]{0pt}{\Tdim}\box0}}

\def\Tr{{\rm Tr}}

\def\p{\partial}

\def\D{\mathcal{D}}

\def\=:{=\hspace{-.7em}\raisebox{1.1ex}{.}\hspace{.1em}\raisebox{-0.2ex}{.} }

\begin{document}

\preprint{RIKEN-MP-13, INT-PUB-11-001}
\title{Confined Monopoles Induced by Quantum Effects in Dense QCD}
\author{Minoru Eto$^1$, Muneto Nitta$^2$, and Naoki Yamamoto$^3$}
\affiliation{
$^{1}$Mathematical physics Laboratory, RIKEN Nishina Center, Saitama 351-0198, Japan\\
$^{2}$Department of Physics, and Research and Education Center for 
Natural Sciences, Keio University, 4-1-1 Hiyoshi, Yokohama, Kanagawa 
223-8521, Japan\\
$^{3}$Institute for Nuclear Theory, University of Washington, Seattle, 
Washington 98195-1550, USA}
\date{\today}

\begin{abstract}
We analytically show that mesonic bound states of confined monopoles 
appear inside a non-Abelian vortex-string in massless three-flavor QCD 
at large quark chemical potential $\mu$.
The orientational modes $\mathbb{C}P^2$ in the internal space of a 
vortex is described by the low-energy effective world-sheet theory. 
Mesons of confined monopoles are dynamically generated as bound states 
of kinks by the quantum effects in the effective theory.
The mass of monopoles is shown to be an exponentially soft scale 
$M \sim \Delta \exp[- c (\mu/\Delta)^2]$,
with the color superconducting gap $\Delta$ and some constant $c$.
A possible quark-monopole duality between the hadron phase 
and the color superconducting phase is also discussed.

\end{abstract}

\pacs{21.65.Qr, 11.27.+d}

\maketitle

\section{Introduction}
Understanding the confinement of quarks and gluons
is one of the most important questions in quantum chromodynamics (QCD).
Although there have been several proposals to explain the origin of the 
confinement, still no consensus has been reached. Among others, one plausible 
scenario is the dual superconducting picture of the QCD vacuum \cite{N74}: 
assuming the condensation of putative magnetic monopoles in the QCD vacuum, 
the color electric flux is squeezed between a quark and an antiquark so that 
the quark-antiquark pair is confined as a meson. This is similar to the situation 
where the magnetic flux is squeezed into a string in the metallic superconductor 
due to the Meissner effect. Although this scenario succeeds in accounting for
a number of properties in the QCD vacuum 
(see, e.g., \cite{SST93, Kondo:1997pc}) 
and is shown to be realized in the ${\cal N}=2$ supersymmetric (SUSY) QCD \cite{SW94}, 
the condensation or even the existence of monopoles cannot be justified 
in real QCD without dramatic assumptions \cite{G03}.

If monopoles indeed exist within the theory of QCD, it is natural to expect
that monopoles would also show up in QCD at finite temperature $T$ 
and finite quark chemical potential $\mu$.
In the quark-gluon plasma phase at high $T$, several instances of 
evidence of the existence of monopoles and their important roles are 
suggested in the model calculations in conjunction with
the lattice QCD simulations (for reviews, see \cite{C08, S08}).
One can question the existence of monopoles in QCD at large $\mu$,
as first addressed by the present authors \cite{ENY10}.
It is indeed an ideal situation to investigate this question at 
asymptotic large $\mu$; the ground state is found to be the most 
symmetric three-flavor color superconductivity  
called the color-flavor locked (CFL) phase \cite{ARW99}
due to the condensation of quark-quark pairing
(for a recent review, see \cite{CSC});
the physics is under theoretical control in this regime
because the QCD coupling constant $g_s$ is weak according to the 
asymptotic freedom.

In the CFL phase, the $\U(1)_B$ symmetry is spontaneously broken by the 
condensation of quark-quark pairing. 
This gives rise to the emergence of Abelian vortices (superfluid vortices)
characterized by the first homotopy group $\pi_1[\U(1)_B] = \mathbb{Z}$ \cite{FZ02}. 
Moreover, owing to the color-flavor locking structure of the pairing, there also 
appear non-Abelian vortices (semi-superfluid vortices) \cite{BDM06} 
having only winding number $1/3$ inside $\U(1)_B$ 
and carrying a color magnetic flux.
The non-Abelian vortices are defined as those characterized by the homotopy group 
$\pi_1(G/H)={\mathbb Z}$ for the symmetry breaking pattern $G \rightarrow H$
with the condition that $H$ is non-Abelian. 
The distinct property of non-Abelian vortices is that they have internal 
collective coordinates (called the orientational modes or the moduli) as a 
consequence of the symmetry breaking in the presence of each vortex. 
In the case of the CFL phase, the moduli of a non-Abelian vortex is the
projective complex space 
$\mathbb{C}P^2 \simeq \SU(3)/[\SU(2)\times \U(1)]$ \cite{NNM08}.
Based on the philosophy of the effective theory (see, e.g., \cite{K95} for a review)
the low-energy effective world-sheet theory for these orientational modes 
near the critical temperature $T_c$ of the CFL phase is constructed in \cite{EN09, ENN09}. 
The interaction between the $\mathbb{C}P^2$ modes in the vortex 
world-sheet and gluons in the bulk has also been determined \cite{Hirono:2010gq}.

Actually, such non-Abelian vortices originating from the color-flavor locking 
were first found in the ${\cal N}=2$ SUSY $\U(N)$ QCD \cite{vortex}.
Remarkably, in the Higgs phase of the ${\cal N}=2$ SUSY QCD, the squark mass leads 
to the {\it dynamical} symmetry breaking pattern $\U(N) \rightarrow \U(1)^N$ 
and supports the existence of monopoles characterized by 
$\pi_2[\U(N)/\U(1)^N]=\mathbb{Z}^N$ \cite{Shifman:2004dr,SY04,SY07}.
In real QCD, on the other hand, it is shown in \cite{ENY10} that 
the strange quark mass $m_s$ together with the charge neutrality and 
$\beta$-equilibrium conditions (required in the realistic dense matter)
exhibits just the {\it explicit} symmetry breaking pattern $\SU(3) \rightarrow \U(1)^2$
and does not support the existence of monopoles dynamically as it should not.

Still there is another mechanism supporting monopoles in real QCD
at large $\mu$ mentioned in \cite{ENY10} in analogy with the SUSY QCD, that is,
the possible nonperturbative quantum fluctuations of the orientational modes.
Such quantum effects are shown to generate a single confined monopole 
attached to non-Abelian vortices in the SUSY QCD \cite{Shifman:2004dr} and 
a monopole-antimonopole meson 
in nonsupersymmetric models motivated by the SUSY \cite{GSY05, GSY06,SY07}.
If this is also the case in real QCD, monopoles must be confined 
due to the color Meissner effect of the color superconductivity in the
Higgs phase.

In this paper, we analytically show that mesonic bound states of
confined monopoles appear as bound states of kinks 
in the effective world-sheet theory on non-Abelian vortices
in massless three-flavor QCD at large $\mu$.
The main difference from our previous analysis in \cite{ENY10} 
is that here we ignore the effect of the strange quark mass $m_s$, 
but take into account the quantum effects of the orientational modes.
In particular, we derive an exponentially soft 
mass scale of confined monopoles near $T_c$:
\beq
\label{eq:result}
M \sim \Delta e^{-c(\mu/\Delta)^2},
\eeq
where $\Delta$ is the superconducting gap and $c$ is some constant.

The existence of mesonic bound states of confined monopoles in the CFL phase
naturally realizes the ``dual" of the putative dual superconducting 
scenario for the quark confinement in the hadron phase.
We also point out the resemblance of the color-octet mesons formed by 
monopole-antimonopole pairs in the CFL phase to
the flavor-octet mesons formed by quark-antiquark pairs in the hadron phase.
This leads us to speculate on the idea of the ``quark-monopole duality,'' i.e., 
the roles played by quarks and monopoles are interchanged between the hadron 
phase and the CFL phase.
This duality, if realized, implies the condensation of monopoles 
in the hadron phase corresponding to the condensation of quarks in the CFL phase,
and thus, embodies the dual superconducting picture in the hadron phase.

The paper is organized as follows. In Sec.~\ref{sec:GL}, we review the time-dependent
Ginzburg-Landau Lagrangian (TDGL). In Sec.~\ref{sec:NA}, we summarize the solution of a 
non-Abelian vortex and the construction of the effective world-sheet theory 
on a non-Abelian vortex. In Sec.~\ref{sec:monopole}, we show that a mesonic 
bound state of confined monopoles appear on a vortex, and discuss its possible 
implications. Section~\ref{sec:discussion} is devoted to conclusion and outlook.

\section{Time-dependent Ginzburg-Landau Lagrangian}
\label{sec:GL}
In this section, we review the TDGL Lagrangian at sufficiently large $\mu$. 
We consider massless three-flavor QCD.
This situation is different from the one considered in \cite{ENY10} 
where the effect of the strange quark mass $m_s$ together with 
the charge neutrality and $\beta$-equilibrium are taken into account.
One may not ignore these effects in the realistic situation, e.g., 
inside the neutron stars.
We will comment on this issue at the end of Sec.~\ref{sec:continuity}.

Let us first introduce the order parameters of the color superconductivity,
the diquark condensates $\Phi_{L,R}$. 
The diquark condensates are induced by the attractive one-gluon exchange 
and the instanton-induced interactions in the color antisymmetric channel 
according to the BCS mechanism \cite{CSC}. 
In Dirac space, the Lorentz scalar (spin-parity $0^+$ channel) 
is the most favorable, since it allows all the quarks near the Fermi 
surface to participate in the pairing coherently.
The positive parity state is favored by the instanton effects \cite{S01}.
The remaining quantum number, the flavor, must be antisymmetrized 
for the pairing to follow the Pauli principle. Therefore, the diquark 
condensate takes the form
\beq
(\Phi_L)_a^i \sim \epsilon_{abc} \epsilon_{ijk}
\langle (q_L)_b^j C (q_L)_c^k \rangle, \nonumber \\
(\Phi_R)_a^i \sim \epsilon_{abc} \epsilon_{ijk}
\langle (q_R)_b^j C (q_R)_c^k \rangle,
\eeq
where $i,j,k$ ($a,b,c$) are flavor (color) indices and $C$ is the charge 
conjugation operator. 
The positive parity ground state is expressed by
\beq
\Phi_L=-\Phi_R=\Phi.
\eeq

We then construct the TDGL Lagrangian based on the QCD symmetry under
\beq
G= \frac{\SU(3)_C \times \SU(3)_F \times \U(1)_B}
{(\mathbb{Z}_3)_{C+B} \times (\mathbb{Z}_3)_{F+B}},
\eeq
where $F$ can be either $L$ or $R$ and 
redundancy of the discrete groups are removed.
Under the symmetry $G$, $\Phi_{F}$ transform as
\beq
\Phi_{F} \rightarrow e^{2i\theta} U_C \Phi_{F} U_{F},
\eeq
where $e^{i \theta} \in \U(1)_B$, $U_C \in \SU(3)_C$, and 
$U_{F} \in \SU(3)_{F}$. 

Because of the absence of the Lorentz invariance in the medium, the 
Lagrangian respects the $\SO(3)$ spatial rotation.
Near the critical temperature $T_c$ of the color superconductivity, 
the order parameters $\Phi_{L,R}$ are sufficiently small
so that higher order terms in $\Phi_{L,R}$ are negligible.
Also, as long as we consider the long-wavelength and low-frequency
deviation from the equilibrium, we can perform the derivative expansion.
Up to the second order in time and space derivatives, the TDGL Lagrangian
invariant under $G$ is given by \cite{GR02, A06}:\footnote{To 
be precise, terms including the first derivative, e.g.,
$\Tr\left(\Phi^{\dagger}\D_0 \Phi \right)$,
related to the dissipation, are not forbidden by the QCD symmetry.
Here we ignore them because they turn out to be irrelevant 
to the dynamics of a non-Abelian vortex finally \cite{ENN09}.}
\beq
{\cal L} \!&=&\! 
\Tr\left(K_0\D_0\Phi^\dagger \D_0\Phi 
- K_3\D_i \Phi^\dagger \D_i \Phi \right) \non & & 
+{\varepsilon\over 2} F_{0i}^2 - {1\over 4 \lambda} F_{ij}^2 
- V, \non
V\!&=&\! \alpha \Tr\left(\Phi^\dagger \Phi \right)
+ \beta_1 \left[\Tr(\Phi^\dagger\Phi)\right]^2 
+\beta_2 \Tr \left[(\Phi^\dagger\Phi)^2\right],
\label{eq:gl}
\eeq
where $\D_\mu \Phi = \p_\mu \Phi - i g_s A_\mu \Phi$, 
$F_{\mu \nu}=\partial_{\mu}A_{\nu}
-\partial_{\nu}A_{\mu}-ig_s[A_{\mu},A_{\nu}]$.
$\varepsilon$ and $\lambda$ are the dielectric constant and the magnetic
permeability, respectively, 
both of which we set unity in our previous works \cite{ENN09, EN09, ENY10}.

The leading-order values of the Ginzburg-Landau (GL) coefficients 
$\alpha$, $\beta_{1,2}$, and $K_{0,3}$ are obtained 
from the weak-coupling calculations at large $\mu$ 
\cite{GR02, A06}:
\beq
\alpha &=& 4 N(\mu) \log \frac{T}{T_c}, \nonumber \\
\beta_1 &=& \beta_2 = \frac{7\zeta(3)}{8(\pi T_c)^2}\, N(\mu)\equiv \beta, \\
K_3 &=& \frac{1}{3}K_0 = \frac{7\zeta(3)}{12(\pi T_c)^2}N(\mu), \nonumber
\label{eq:weak}
\eeq
where $N(\mu) = {\mu^2}/({2\pi^2})$ is the density of state at the Fermi 
surface and $T_c = 2^{1/3}e^\gamma \Delta/{\pi}$ is the critical 
temperature of the CFL phase \cite{S02}. These GL coefficients
can be derived by generalizing the computations known in nonrelativistic 
systems \cite{AT66}.

Using the GL potential $V_{\rm GL}$ with the GL coefficients in Eq.~(\ref{eq:weak}), 
one finds the most stable ground state,
\beq
\label{eq:ground}
\Phi = {\rm diag}(\Delta, \Delta, \Delta),
\eeq
where $\Delta=\sqrt{-\alpha/(8\beta)}$. This form of the
ground state entangles the color and flavor rotations and 
is called the color-flavor locked phase.
In the CFL phase, the symmetry $G$ is broken down to
\beq
H=\frac{\SU(3)_{C+F}}{(\mathbb{Z}_3)_{C+F}},
\eeq
and the order parameter manifold is
\beq
G/H \simeq \frac{\SU(3)_{C-F}\times \U(1)_B}{(\mathbb{Z}_3)_{C-F+B}}=\U(3).
\eeq
This is parametrized by would-be Nambu-Goldstone (NG) modes
associated with the dynamical symmetry breaking of $\SU(3)$,
which are eaten by the eight gluons by the Higgs mechanism, 
and a massless NG mode (referred to as the $H$ boson) associated with 
the symmetry breaking of $\U(1)_B$.

By expanding $\Phi$ from the ground state (\ref{eq:ground}),  
\beq
\Phi=\Delta{\bf 1}_3 + \frac{\phi_1 + i \varphi}{\sqrt{2}}{\bf 1}_3
+ \frac{\phi_8^a + i \zeta^a}{\sqrt{2}}T^a,
\eeq
where $\phi_1$ and $\phi_8^a$ ($\varphi$ and $\zeta^a$) 
are real (imaginary) parts of fluctuations, 
mass spectra are obtained as
\beq
\label{eq:spectra0}
m_G^2 = 2\lambda g_s^2 \Delta^2 K_3, \quad
m_{1}^2 = - \frac{2\alpha}{K_3}, \quad
m_{8}^2 = \frac{4\beta \Delta^2}{K_3}.
\eeq
Here $m_G$ is the mass of the gluons 
which absorb $\zeta^a$ by the Higgs mechanism, 
and $m_1$ and $m_8$ are the masses of 
$\phi_1$ and $\phi_8^a$ 
in the ${\bf 3} \otimes {\bf 3}^* = {\bf 1} \oplus {\bf 8}$ 
representation under the 
unbroken $\SU(3)_{C+F}$ symmetry, respectively.

From Eqs.~(\ref{eq:weak}) and (\ref{eq:spectra0}), we have 
\beq
\label{eq:spectra}
m_G \sim g_s \mu, \quad m_1 = 2m_8 \sim 2\Delta.
\eeq 
Because $g_s \mu \gg \Delta$ at large $\mu$, 
the CFL phase is a type-I superconductor as indicated by the 
Ginzburg-Landau parameters \cite{GR03}:
\beq
\kappa_{1,8}=\frac{m_{1,8}}{m_G} \ll 1,
\eeq
where we define two GL parameters corresponding to two coherence lengths
$1/m_{1,8}$.
Note that non-Abelian vortices can appear even in this 
type-I system, since their interactions are repulsive at large distances 
due to the exchange of the $H$ boson \cite{NNM08}. This is in contrast to the case 
of the metallic (Abelian) superconductor where vortices can appear only in 
a type-II system with $\kappa>1$ (under a suitable normalization). 
Non-Abelian vortices are the superfluid vortices and 
are created under a rapid rotation.

\section{Non-Abelian vortices}
\label{sec:NA}
In this section, we consider the properties of a non-Abelian vortex 
in the CFL phase for later discussions in Sec.~\ref{sec:monopole}. 
The results are already obtained in our previous papers \cite{ENN09, EN09, ENY10}.
We also correct some of equations given in these references.

\subsection{Non-Abelian vortex solutions}
We first consider a non-Abelian vortex solution in the CFL phase.
We make the standard ansatz for a static vortex-string configuration 
parallel to the $x_3$ direction 
(perpendicular to the $x_1$-$x_2$ plane):
\beq
\Phi(r,\theta) &=& \Delta \, {\rm diag}\left(
e^{i\theta}f(r),\ g(r),\ g(r)\right),\\
A_i(r,\theta) &=& \frac{1}{g_s} \frac{\epsilon_{ij}x^{j}}{r^2}
 \left[1-h(r)\right] 
{\rm diag}
\left(
-\frac{2}{3},\ \frac{1}{3},\ \frac{1}{3}
\right),\non 
\label{eq:minimum-winding}
\eeq
with $i,j=1,2$. 
This ansatz can be rewritten as
\beq
\Phi &=& \Delta \,
e^{i\theta\left(\frac{1}{\sqrt{3}}T_0-\sqrt{\frac{2}{3}}T_8 \right)}
\left(
\frac{F}{\sqrt 3} T_0 - \sqrt{2 \over 3} G T_8
\right), \label{eq:ans_s} 
\\
A_i &=& \frac{1}{g_s} \frac{\epsilon_{ij}x^{j}}{r^2}
 \left(1-h\right) \sqrt{\frac{2}{3}} T_8 ,
\label{eq:ans_a}
\eeq 
with profile functions 
\beq
F \equiv f + 2g,\quad G \equiv f-g,
\eeq
and the $\U(3)$ generators
\beq
 T_0 \equiv \frac{1}{\sqrt{3}}{\rm diag}(1,1,1),\quad
 T_8 \equiv \frac{1}{\sqrt{6}}
 {\rm diag}(-2,1,1) .
\eeq

Equations of motion for the profile functions $f$, $g$, and $h$ are of the form
\beq
\left[ \triangle 
-\frac{(2 h+1)^2}{9 r^2}-\frac{m_1^2}{6}  \left(A-3\right) -\frac{m_8^2}{3} B\right]f = 0,
\label{eq:1}\\
\left[ \triangle 
-\frac{(h-1)^2}{9 r^2}-\frac{m_1^2}{6}  \left(A-3\right) 
+\frac{m_8^2}{6}  B \right]g =0,
\label{eq:2}\\
h''-\frac{h'}{r} - \frac{m_G^2}{3}  \left(g^2 (h-1)+f^2 (2 h+1)\right) = 0,
\label{eq:3}
\eeq
with the Laplacian $\triangle \equiv \p_r^2 + r^{-1}\p_r$, 
$A\equiv f^2 + 2g^2$ and $B \equiv f^2 - g^2$.
These differential equations should be solved 
with the boundary conditions
\beq
\left\{
\begin{array}{cl}
(f,g,h) \to (1,1,0) \quad &\text{as}\quad r \to \infty,\\
(f,g',h) \to (0,0,1) \quad &\text{as}\quad r \to 0.
\end{array}
\right.
\label{eq:bc}
\eeq
Numerical solutions can be found in \cite{EN09, ENN09}.

\subsection{Effective world-sheet theory on a vortex}
In this subsection we construct the effective world-sheet 
theory on a single non-Abelian vortex, 
which describes fluctuations of the ${\mathbb C}P^2$ 
orientational modes $\phi$. 
We place the vortex-string along the $x_3$ axis,
so we construct the effective action in the $(t,x_3)$ coordinates
by integrating over the $x_1$-$x_2$ plane.

First of all, we take a singular gauge 
in which the single vortex configuration is expressed as
\beq
\Phi^{\star} &=& \Delta \,
e^{\frac{i\theta}{3}}
\left(
\frac{F(r)}{\sqrt 3} T_0 - \sqrt{2 \over 3} G(r) T_8
\right),
\\
A_i^\star &=& - \frac{1}{g_s} \frac{\epsilon_{ij}x^{j}}{r^2}
h(r) \sqrt{\frac{2}{3}} T_8. 
\eeq
Then the general solution can be reproduced 
by acting the color-flavor locked symmetry on them:  
\beq
  \Phi(U) \to U \Phi^\star U^{-1},\quad
   A_i(U) \to U A_i^\star U^{-1},
\eeq
where $U \in \SU(3)_{\rm C+F}$.
This action changes only $T_8$ 
with $T_0\sim {\bf 1}_3$ unchanged. 
We define coordinates $\phi$ on $\mathbb{C}P^2$ by 
\beq
- U \left(\sqrt{\frac{2}{3}}T_8\right) U^{-1} \equiv   \phi  \phi^\dagger - \frac{{\bf 1}_3}{3}
\equiv \left< \phi \phi^\dagger\right>, \label{eq:phi}
\eeq
where $\phi$ is a complex $3$-column vector, 
and $\left< {\cal O} \right>$ denotes the traceless part 
of a square matrix ${\cal O}$. 
The $\SU(3)_{\rm C+F}$ symmetry acts on $  \phi$ from the left hand side as $  \phi \to U   \phi$.
Taking trace of this gives a constraint 
\beq
\label{eq:constraint}
 \phi^\dagger   \phi = 1.
\eeq
Since the phase of $\phi$ is redundant, 
we find that $\phi$ represents the homogeneous coordinates 
of the complex projective space ${\mathbb C}P^2$.

Physically, these degrees of freedom (called the orientational modes
or the moduli) arise associated with the symmetry breaking from 
the $\SU(3)_{C+F}$ symmetry preserved by the diquark condensates 
to $[\U(1)\times \SU(2)]_{C+F}$ in the presence of each vortex
\beq
\frac{\SU(3)_{C+F}}{[\U(1) \times \SU(2)]_{C+F}} \simeq 
\mathbb{C}P^2.
\eeq
The NG modes $\phi \in \mathbb{C}P^2$ 
propagate along the non-Abelian vortex-string.
The form of the Lagrangian is determined solely by the 
$\SU(3)/[\U(1)\times \SU(2)]$ symmetry, 
and is described by the $\mathbb{C}P^2$ nonlinear sigma model.

The effective Lagrangian consists of two parts:
\beq
\! \! \! {\cal L}_{\rm eff} 
\!&=&\! 
{\cal L}_{\rm eff}^{(0)} + {\cal L}_{\rm eff}^{(3)}  \label{eq:Lag-dec}\\ 
{\cal L}_{\rm eff}^{(0)} \!\!&=&\!\!\! 
\int\! dx_1dx_2 \Tr \! \left[-{\varepsilon\over 2} F_{0i}F^{0i} \! + \! 
K_0 {\cal D}_0 \Phi^\dagger {\cal D}^0\Phi \right] \!, \\
{\cal L}_{\rm eff}^{(3)} \!\!&=&\!\!\! 
\int\! dx_1dx_2 \Tr \! \left[- {1\over 2\lambda} F_{3i}F^{3i} \! 
+ \! K_3 {\cal D}_3 \Phi^\dagger {\cal D}^3\Phi \right] \!,
\eeq
where the ${\mathbb C}P^2$ 
orientational modes $\phi$ are promoted 
to fields $\phi(t,x^3)$ on the vortex world sheet. 
For gauge field we use the ansatz of Gorsky-Shifman-Yung \cite{GSY05}:
\beq
 A_{\alpha}(\phi (x^{\alpha})) \equiv {i \rho_{\alpha}(r)\over g_s} 
[\left<\phi \phi^\dagger\right>,\partial_{\alpha}\left<\phi \phi^\dagger\right>], \quad \alpha =0,3, 
\eeq
where functions $\rho_{\alpha}(r)$ ($\alpha=0,3$) 
are undetermined at this stage, 
and will be determined below. 
Note that we need two independent functions in this ansatz 
due to the absence of the Lorentz invariance in medium unlike \cite{GSY05}, 
which was not considered in our previous paper \cite{ENN09}.

For later use, it is convenient to define the function ${\cal F}_\alpha$  by \cite{ENN09}
\beq
 {\cal F}_{\alpha}(a,b) \equiv a \phi \partial_{\alpha}\phi^\dagger  
+ b \partial_{\alpha}\phi \phi^\dagger  + (a-b) \phi\phi^\dagger \partial_{\alpha}\phi \phi^\dagger, 
\eeq
with $a,b \in {\mathbb C}$, which satisfies 
\beq
&& \Tr \left[{\cal F}^{\alpha}(a,b)^\dagger {\cal F}^{\alpha}(a,b)\right] 
 = (|a|^2+|b|^2) {\cal L}_{{\mathbb C}P^2} .
\eeq
Here ${\cal L}_{{\mathbb C}P^2}$ is the form of 
the ${\mathbb C}P^2$ nonlinear sigma model Lagrangian: 
\beq
&&{\cal L}_{{\mathbb C}P^2} \equiv 
  \partial^{\alpha}\phi^\dagger \partial_{\alpha}\phi
 + (\phi^\dagger \partial_{\alpha}\phi)(\phi^\dagger \partial^{\alpha}\phi).
\eeq

By using the function ${\cal F}_\alpha$, parts of the Lagrangian can be rewritten as
\beq
 {\cal D}_{\alpha} \Phi &=& \Delta e^{i \theta/3} {\cal F}_{\alpha}(f-g+\rho_{\alpha}g,f-g-\rho_{\alpha}f),  \\
 F_{\alpha i}&=& {1\over g_s} \epsilon_{ij} {x_j \over r^2} h(1-\rho_{\alpha}){\cal F}_{\alpha}(1,1) 
 - {i \over g_s} {x_j \over r} \rho_{\alpha}^\prime {\cal F}_{\alpha}(1,-1). \non
\eeq
By using these expressions, we calculate each term in 
the Lagrangian. 
First, the term of the gauge field strength can 
be calculated as
\begin{widetext}
\beq
\Tr \left[F_{i\alpha} F^{i\alpha} \right]
 &=& {1\over g_s^2} {x_i x^i\over r^4} h^2 g(1-\rho_{\alpha})^2 
{\cal G}_\alpha \left(1,1|1,1\right)
  - {2i \over g_s^2} {\epsilon_{ij}x^i x^j \over r^3} 
  h (1-\rho_{\alpha}) \rho_{\alpha}^\prime  
{\cal G}_\alpha \left(1,1|1,-1\right)
- {1\over\ g_s^2} {x_i x^i \over r^2} (\rho_{\alpha}^\prime)^2 
{\cal G}_\alpha \left(1,-1|1,-1\right) \non
&=& - {2\over g_s^2} \left[(\rho_{\alpha}^\prime)^2 
      + {h^2(1-\rho_{\alpha})^2\over r^2} \right] 
        {\cal L}^{(\alpha)}_{{\mathbb C}P^2}  ,\label{eq:FF}
\eeq
\end{widetext}
with ${\cal G}_\alpha \left(k,l|m,n\right) \equiv \Tr  [{\cal F}_{\alpha}(k,l){\cal F}^{\alpha}(m,n)]$
and no summation is taken for $\alpha$.
Here we have used the following relations 
\beq
{\cal G}_\alpha \left(1,1|1,1\right) = - {\cal G}_\alpha \left(1,-1|1,-1\right)
= 2  {\cal L}^{(\alpha)}_{{\mathbb C}P^2} .
\eeq
Similarly the term including $\Phi$ can be calculated to give
\beq
\Tr \left[{\cal D}_{\alpha}\Phi^\dagger {\cal D}^{\alpha}\Phi\right]
&=&2\Delta^2 \bigg[(1-\rho_{\alpha})(f-g)^2 \non
&&+ {\rho_{\alpha}^2\over 2}(f^2+g^2)\bigg] {\cal L}_{{\mathbb C}P^2}^{(\alpha)}.
 \label{eq:DPhi2}
\eeq

Substituting Eqs.~(\ref{eq:FF}) and (\ref{eq:DPhi2}) into (\ref{eq:Lag-dec}), 
we finally obtain the ${\mathbb C}P^2$ Lagrangian\footnote{The coefficients
$CK_{0,3}$ in Eq.~(3.18) of Ref.~\cite{ENN09} should be corrected as $C_{0,3}$.}
\beq
  {\cal L}_{\rm eff} 
 = C_0 {\cal L}_{{\mathbb C}P^2}^{(0)} 
 + C_3 {\cal L}_{{\mathbb C}P^2}^{(3)} \label{eq:eff-Lag},
\eeq
with two different coefficients for time and space components, 
\beq
\label{eq:Kahrer-int1}
\! \! \! C_0 \! &=& \! {4\pi \over \lambda g_s^2} 
 \int \! dr {r\over 2}\left[\varepsilon\lambda \left((\rho_0^\prime)^2 + {h^2\over r^2}(1-\rho_0)^2\right) \right. \non
&&\left. 
\! \! + {K_0 \over K_3} m_G^2 \left(\!(1-\rho_0)(f-g)^2 
  + {f^2+g^2 \over 2}\rho_0^2 \right)\right] \!, \\
\label{eq:Kahrer-int2}
\! \! \! C_3 \! &=& \! {4\pi \over \lambda g_s^2} \int \! dr {r\over 2}\left[
(\rho_3^\prime)^2 + {h^2\over r^2}(1-\rho_3)^2 \right. \non
&&\left. 
 + m_G^2 \left(\!(1-\rho_3)(f-g)^2 + {f^2+g^2 \over 2}\rho_3^2 \right)\right] \!.
\eeq
$C_0$ and $C_3$ should be determined by minimizing them 
through $\rho_{0,3}$:
\beq
&& \rho_0^{\prime\prime} + {\rho_0^\prime \over r} 
 + (1-\rho_0) {h^2 \over r^2} \non
&& - {K_0 \over \varepsilon \lambda K_3} m_G^2 \left[ (f^2+g^2)\rho_0 - (f-g)^2\right] 
=0,\\
&& \rho_3^{\prime\prime} + {\rho_3^\prime \over r} 
 + (1-\rho_3) {h^2 \over r^2} \non
&& - m_G^2 \left[ (f^2+g^2)\rho_3 - (f-g)^2\right] =0.
\eeq
From Eqs.~(\ref{eq:spectra}), (\ref{eq:Kahrer-int1}),
and (\ref{eq:Kahrer-int2}), $C_{0,3}$ are estimated as
\beq
\label{eq:Kahrer}
C_{0,3} \sim \frac{\mu^2}{\Delta^2}.
\eeq

The velocity of the ${\mathbb C}P^2$ modes propagating 
along the vortex-string is then 
\beq
 v^2 = C_3 / C_0 .
\eeq
This nontrivially depends on $\mu$, 
which we do not discuss in detail in this paper. 
By rescaling
\beq
\label{eq:rescale}
t\to \sqrt{C_0} t, \qquad x^3 \to \sqrt{C_3} x^3,
\eeq 
the Lagrangian (\ref{eq:eff-Lag}) is cast in 
the Lorentz invariant form
\beq
\label{eq:CP2}
 {\cal L}_{\rm eff} = {\cal L}_{{\mathbb C}P^2} =
  \partial^{\alpha}\phi^\dagger \partial_{\alpha}\phi
 + (\phi^\dagger \partial_{\alpha}\phi)(\phi^\dagger \partial^{\alpha}\phi).
\eeq
Note that the first derivative terms ignored
in Eq.~(\ref{eq:gl}) do not contribute to the effective theory
because of the property ${\rm tr}({\cal F}_\alpha)=0$ \cite{ENN09}.

\section{Confined monopoles}
\label{sec:monopole}
In this section, we show that mesonic bound states of confined
monopoles appear inside the non-Abelian vortices by solving 
the effective world-sheet theory constructed in the previous section
in the large-$N_c$ limit.
We further argue a possible ``quark-monopole duality" between the
hadron phase and the color superconducting phase.

\subsection{Bound state of monopole-antimonopole: 
Kink-antikink pairing on a vortex}
\label{sec:solve}
In this subsection, we consider the properties of the solution to the 
${\mathbb C}P^2$ nonlinear sigma model Lagrangian for the orientational 
modes, by taking into account the quantum effects. Thereby
we will find that there appear a kink-antikink pairing on a vortex
which can be identified as the mesonic bound state of a monopole and an 
antimonopole in the 3+1 dimensions.

Unfortunately, however, the solution to the $\mathbb{C}P^2$ nonlinear sigma model 
is not known so far, 
although the $\mathbb{C}P^1$ model 
[equivalent to the ${\rm O}(3)$ nonlinear sigma model]
is solved rigorously \cite{ZZ78}. 
Here we consider the $\mathbb{C}P^{N_c-1}$ model instead and solve
the model to leading order of $1/N_c$ following \cite{DLD78, W78}.
Owing to the qualitative similarity of the solutions to the $\mathbb{C}P^1$ 
and $\mathbb{C}P^{N_c-1}$ models, the solution to the
$\mathbb{C}P^2$ model should be approximately described 
by the solution to the $\mathbb{C}P^{N_c-1}$ model with taking $N_c=3$ at the end.
This is the only assumption which we will make in our calculations.
To make our paper self-contained, we shall describe the original 
derivation \cite{DLD78, W78} for our Lagrangian (\ref{eq:CP2}) under the constraint 
(\ref{eq:constraint}) in the following.

We first perform the Hubbard-Stratonovich transformation by introducing 
the auxiliary field to eliminate the quartic term in Eq.~(\ref{eq:CP2}).
Because the quartic term is the vector-vector type interaction, the 
auxiliary field should be the gauge field $A_{\alpha}$. 
After the Hubbard-Stratonovich transformation, the Lagrangian becomes
\beq
\label{eq:aux}
{\cal L}_{\rm eff}= 
(\partial_{\alpha}-iA_{\alpha}) \phi^\dagger (\partial^{\alpha}+iA^{\alpha}) \phi, 
\eeq
with the constraint (\ref{eq:constraint}).
Actually, eliminating $A_{\alpha}$ by using the equation of motion for $A_{\alpha}$,
one can easily check that Eq.~(\ref{eq:aux}) reduces to Eq.~(\ref{eq:CP2}).
Equation (\ref{eq:aux}) can be regarded as the $\U(1)_D$ gauge theory 
(``D" stands for a dummy gauge symmetry);
it has the local gauge symmetry $\phi \rightarrow e^{i\theta(x)} \phi$ 
under the
gauge transformation $A_{\alpha} \rightarrow A_{\alpha} - \partial_{\alpha}\theta(x)$.
To take into account the constraint (\ref{eq:constraint}) in the Lagrangian,
we then introduce another auxiliary field $\sigma$ as a Lagrange multiplier,
to obtain
\beq
\label{eq:aux2}
{\cal L}_{\rm eff}= 
(\partial_{\alpha}-iA_{\alpha}) \phi^\dagger (\partial^{\alpha}+iA^{\alpha}) \phi 
- \sigma \left(\phi^{\dag}\phi - 1 \right).
\eeq
This expression of the ${\mathbb C}P^{N_c-1}$ model is nothing but 
the K\"ahler quotient.
After rescaling the $\phi$ and $\phi^{\dag}$ variables,
\beq
\label{eq:rescale2}
\phi \rightarrow \left( \frac{1}{C_0 C_3} \right)^{\! \! 1/4} \! \! \phi, \quad
\phi^{\dag} \rightarrow \left( \frac{1}{C_0 C_3} \right)^{\! \! 1/4} 
\! \! \phi^{\dag},
\eeq
the partition function of the theory is given by
\beq
Z=\int [d\phi d\phi^{\dag} d\sigma dA_{\alpha}] \ e^{iS},
\eeq
where the action is given by
\beq
\label{eq:action}
S&=&\int dx_0 dx_3 \biggl[(\partial_{\alpha}-iA_{\alpha}) \phi^\dagger 
(\partial^{\alpha}+iA^{\alpha}) \phi 
\non
& & \qquad \qquad \quad - \sigma 
\left(\phi^{\dag}\phi -  \frac{N_c}{3}\sqrt{C_0 C_3} \right) \biggr].
\eeq
Note here that the coefficient of the kinetic term becomes unity
due to the rescaling of the fields, Eq.~(\ref{eq:rescale2}), together with 
the rescaling of measure, Eq.~(\ref{eq:rescale}).
The prefactor of $N_c/3$ in Eq.~(\ref{eq:action}) is introduced 
to perform the $1/N_c$ expansion consistently below, and is chosen such that
it reduces to unity for $N_c=3$.

Integrating out $\phi$ and $\phi^{\dag}$, one obtains
\beq 
\! \! \! \! \! \! \! \! \! \! \! \! \! \! \! 
Z &=& \int [d\sigma dA_{\alpha}] \exp
\left[-N_c {\rm tr} \ln \biggl(-(\partial_{\alpha}+iA_{\alpha})^2 
- \sigma \biggr)
\right. 
\non
& & \qquad \qquad \qquad \ \ 
+ i \frac{N_c}{3}\sqrt{C_0 C_3} \int dx_0 dx_3 \ \sigma 
\biggr],
\label{eq:Z}
\eeq
The Lorentz invariance implies the saddle point $A_{\alpha}=0$ and 
constant $\sigma$.

To leading order of $1/N_c$,
varying the partition function with respect to $\sigma$ 
gives the gap equation:
\beq
i \frac{\sqrt{C_0 C_3}}{3}  + \int^{\Lambda=\Delta} \frac{d^2k}{(2\pi)^2}
\frac{1}{k^2-\sigma + i\epsilon}=0, \non
\eeq
where we have introduced the cutoff of the low-energy effective theory,
$\Lambda = \Delta$.
This is not a dynamical cutoff of the ${\mathbb C}P^2$ model
but is a physical cutoff, namely $\Delta$ is the mass gap of 
the quasiparticles of quarks in the original GL Lagrangian [see Eq.~(\ref{eq:spectra})].
After the integral, one arrives at
\beq
\label{eq:saddle}
M^2 \equiv \sigma \sim \Delta^2 e^{-4\pi \sqrt{C_0 C_3}/3}.
\eeq
Using (\ref{eq:Kahrer}), $M$ is expressed as
\beq
\label{eq:mass}
M \sim \Delta e^{-c(\mu/\Delta)^2},
\eeq
with some constant $c$.
$M$ can be now identified as the mass of $\phi$ 
and $\phi^{\dag}$ induced by the quantum effects from Eq.~(\ref{eq:action}).
This mass gap for the orientational modes is required 
by the Coleman-Mermin-Wagner theorem in the 1+1 dimensions,
as mentioned in \cite{ENY10}.

We then consider the fluctuations around the saddle
point $A_{\alpha}=0$ and $\sigma$ given in Eq.~(\ref{eq:saddle}).
For this purpose, we expand the partition function with respect to
$\sigma$ and $A_\alpha$. Since higher order terms in $\sigma$ and $A_\alpha$ are
suppressed in the large-$N_c$ limit, only the quadratic terms are relevant in the 
following. Expansion of the functional determinant in Eq.~(\ref{eq:Z}) can be
understood in terms of the Feynman diagrams. It turns out that the relevant diagrams 
are the propagator of $A_\alpha$ at one-loop level 
[the $\U(1)_D$ ``photon" self-energy].
Finally, the dynamically generated kinetic term of the gauge field
reduces to $c N_c (-g_{\mu\nu}k^2 + k_\mu k_\nu)$ 
with the coefficient $c = 1/(12\pi M^2)$ \cite{W78}.
The form of the kinetic term is fixed by the gauge invariance.

Now the effective world-sheet theory including 
the quantum effects to leading order of 
$1/N_c$ is summarized as
\beq
{\cal L}_{\rm eff}^{\rm quant} &=& 
(\partial_{\alpha}-iA_{\alpha}) \phi^\dagger (\partial^{\alpha}+iA^{\alpha}) \phi
- M^2 \phi^{\dag}\phi 
\non
& & - \frac{N_c}{48\pi M^2}F_{\alpha \beta}^2.
\eeq
By rescaling $A_{\alpha}$ so that the kinetic term of $A_{\alpha}$ is canonically
normalized,
\beq
\label{eq:rescale3}
A_{\alpha} \rightarrow \sqrt{\frac{12\pi M^2}{N_c}} A_{\alpha},
\eeq
the effective Lagrangian reduces to
\beq
\label{eq:EFT-quantum}
{\cal L}_{\rm eff}^{\rm quant} &=& 
\left(\partial_{\alpha}-iA_{\alpha} M \sqrt{\frac{12\pi}{N_c}}\right) \phi^\dagger
\left(\partial^{\alpha}+iA^{\alpha} M \sqrt{\frac{12\pi}{N_c}}\right) \phi 
\non
& &
- M^2 \phi^{\dag}\phi - \frac{1}{4} F_{\alpha \beta}^2.
\eeq
This implies that $\phi$ and $\phi^{\dag}$ have the effective $\U(1)_D$ 
charges $\pm M \sqrt{12\pi/N_c}$.
Since we are considering the 1+1 dimensions, 
$\phi$ and $\phi^{\dag}$ are confined by the linear potential
\beq
V(x,y)=\frac{12\pi M^2}{N_c}|x-y|,
\eeq
where $x$ and $y$ are the $x^3$ coordinates of $\phi$ and $\phi^{\dag}$.

We are now ready to understand the confining potential between $\phi$
and $\phi^\dagger$ from the 3+1 dimensional viewpoint.
Remembering that $\phi$ is the orientational moduli of the non-Abelian vortex,
a quantum state of $\mathbb{C}P^{N_c-1}$ model is in one-to-one correspondence 
with a quantum vortex state.
Since there exists only one ground state in the $\mathbb{C}P^{N_c-1}$ model,
so is the quantum vortex state whose orientation is not
fixed in a particular direction.
When $\phi$ and $\phi^\dagger$ are placed at positions $x$ and $y$, 
because of the linear potential between them,
the string tension of the vortex 
between $x$ and $y$ is larger than that outside this region by
$\sim M^2/N_c$;
the inner vortex is an excited state compared with the outer vortices (ground state).

\begin{center}
\begin{figure}[h]
\includegraphics[width=0.3 \textwidth] {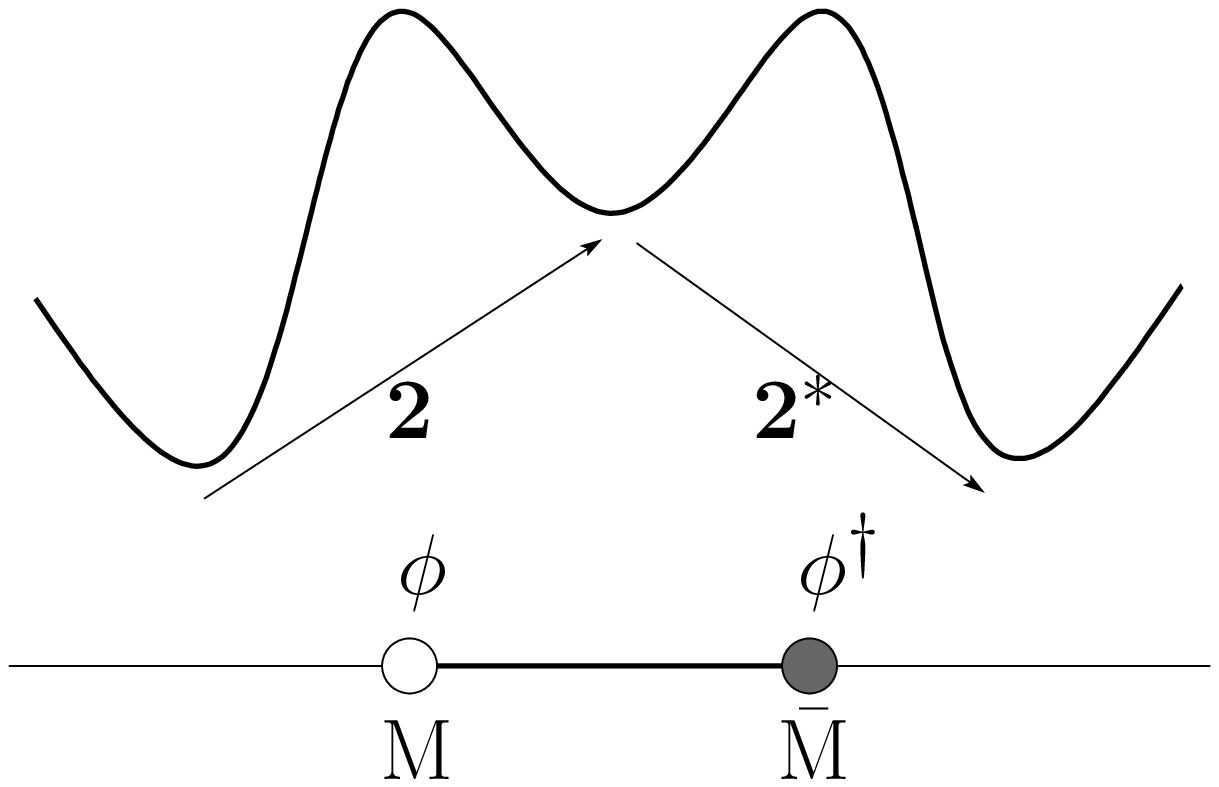} 
\\ (a) \\ \vspace{1cm}
\includegraphics[width=0.3 \textwidth] {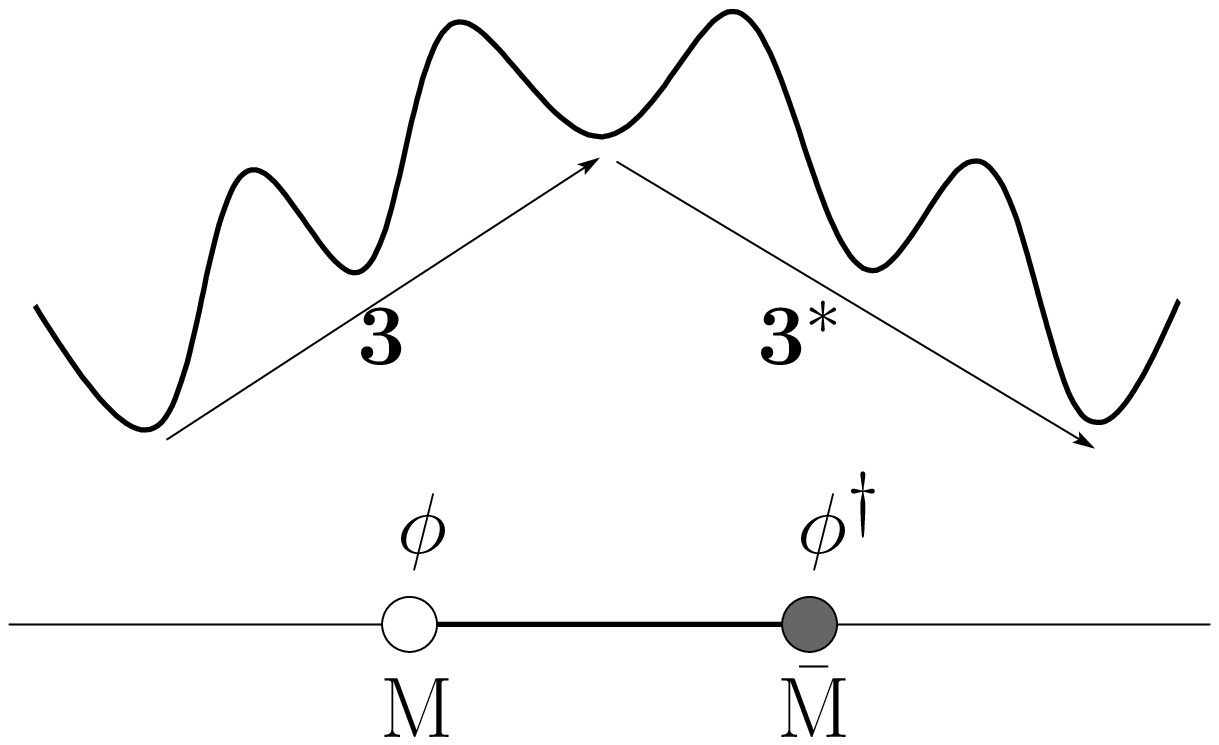}
\\ (b) 
\caption{A schematic illustration of 
the nonperturbative potential and kinks interpolating between 
the ground state and the metastable states, 
in the cases of (a) $N_c=2$ and (b) $N_c=3$.
Kinks ($\phi$ and $\phi^{\dag}$) can be identified with monopoles 
($M$ and $\bar M$) from the bulk $3+1$ dimensional point of view. 
The total configuration is a bound state of a monopole and 
an antimonopole.
}
\label{fig:monopole-pair}
\end{figure}
\end{center}

There is another perspective for understanding of this phenomenon \cite{GSY05}.
The vacuum structure of the $\mathbb{C}P^{N_c-1}$ model can be realized by
looking at the $\theta$-dependence of the theory \cite{Witten:1998uka,Shifman:1998if,GSY05}
\beq
{\cal L}_{\theta}=\frac{\theta}{2\pi}M\sqrt{\frac{12\pi}{N_c}}\epsilon^{\alpha \beta}
\partial_{\alpha}A_{\beta},
\eeq
where the gauge field $A_{\alpha}$ is the one after rescaling (\ref{eq:rescale3}). 
Recalling that the vacuum energy $E(\theta)$ is of order $N_c$ 
in the large-$N_c$ limit, $E(\theta)$ is expressed as 
\beq
E(\theta) = N_c f\left(\frac{\theta}{N_c} \right).
\eeq
Here $f(\theta)$ is an even function of $\theta$ 
due to the $CP$ symmetry under which $\theta$ transforms as 
$\theta \rightarrow -\theta$.
$E(\theta)$ must also satisfy the periodicity,
\beq
E(\theta)= E(\theta+2\pi).
\eeq
One might suspect that these two conditions are incompatible at first sight.
However, there is a way out: both of them can be satisfied
when $E(\theta)$ is a multibranched function as
\beq
\label{eq:theta}
E(\theta)= N_c \min_k f\left(\frac{\theta + 2\pi k}{N_c} \right),\ (k=0,1,\cdots,N_c-1).
\nonumber \\
\eeq  
Expanding $f(\theta)= f(0) + (1/2) f''(0) \theta^2 + \cdots$ 
and considering that higher order terms in $\theta$ are suppressed at large $N_c$
in Eq.~(\ref{eq:theta}), the vacuum energy at $\theta=0$ is given by
\beq
E(0) = E_0 + C \frac{M^2}{N_c}k^2,\ (k=0,1,\cdots,N_c-1),
\eeq
with some constant $C$.
Therefore, there exist $N_c$ local minima among which only one is a true ground state 
while the others are quasivacua, see Fig.~\ref{fig:monopole-pair}. 

It is now natural to interpret $\phi$ and $\phi^\dagger$
as a kink and an antikink interpolating the adjacent local minima on a vortex, 
respectively  \cite{GSY05}; taking into account the codimension, 
this bound state neutral to the $\U(1)_D$ charge can be identified 
as the bound state of a monopole and an antimonopole in terms of the original 
3+1 dimensions, as illustrated in Fig.~\ref{fig:monopole-pair}: 
a monopole and an antimonopole with the mass $M$ 
are confined into the mesonic bound state by the linear potential.
A similar understanding has been demonstrated in \cite{Markov:2004mj}
based on the comparison with the SUSY QCD.

Let us discuss the representations of these objects.
The fields $\phi$ and $\phi^{\dag}$ in the effective theory 
transform as ${\bf N_c}$ and ${\bf N_c}^*$ (anti)fundamental
representations under $\SU(N_c)$, respectively.
The degrees of freedom of $\phi$ is $2(N_c-1)$ after fixing 
$\U(1)_D$ gauge symmetry.
For instance $\phi = {1 \over \sqrt {1+|b_1|^2}}(1,b_1)^T$ 
for $N_c=2$, and 
$\phi = {1 \over \sqrt {1+|b_1|^2 + |b_2|^2}}(1,b_1,b_2)^T$ 
for $N_c=3$. 
For $N_c=2$, $\phi$ ($\phi^{\dag}$) represents one (anti)kink,
as can be seen in Fig.~\ref{fig:monopole-pair}(a).
Each of them corresponds to one (anti)monopole in the bulk.
For $N_c=3$, which is the case of the CFL phase, 
one (anti)monopole is a composite state of $N_c-1=2$ (anti)kinks, 
each of which has one complex moduli (position and phase), 
as seen in Fig.~\ref{fig:monopole-pair}(b). 

We conclude that (anti)monopoles belong to ${\bf N_c}$ (${\bf N_c}^*$) 
fundamental representations of $\SU(N_c)$, 
and they appear as a mesonic bound state.
This mesonic bound state belong to 
${\bf N_c} \otimes {\bf N_c}^* = {\bf 1} \oplus {\bf N_c^2-1}$ 
representation. It was shown in \cite{ZZ78} that 
the singlet in this decomposition does not appear in the spectrum 
 in the ${\mathbb C}P^1$ model ($N_c=2$).
This was interpreted in \cite{Markov:2004mj} that 
the singlet corresponds to a set of monopole and antimonopole 
with opposite charges, which is unstable to decay.
Although there is no such a calculation for $N_c \geq 3$, 
we expect that the same holds.

Before closing this subsection, let us make one comment on fermions.
As discussed in \cite{W78, D'Adda:1978kp}, fermions can be incorporated 
to the ${\mathbb C}P^{N_c-1}$ model. 
In fact, quasiparticles of quarks are shown to be trapped 
in the core of non-Abelian vortices in \cite{Yasui:2010yw},
in which fermion zero modes belonging to 
the triplet of the $\SU(2)$ unbroken symmetry in the core of the vortex 
have been found. However coupling to the bosonic ${\mathbb C}P^2$ model 
is not known yet.

In summary, we found a mesonic bound state of confined monopoles 
with the mass $M$ given in Eq.~(\ref{eq:mass}) induced by the 
quantum effects inside a non-Abelian vortex.

\subsection{A possible quark-monopole duality}
\label{sec:continuity}
\begin{table*}[t]
\begin{center}
  \begin{tabular}{|l c c|}
    \hline
    Phases & Hadron phase (hyper nuclear matter) & Color-flavor locked phase \\ \hline \hline
          & Confinement   & Higgs \\
    Quarks & Confined & Condensed \\ 
    {\it Monopoles}  & {\it Condensed?} & {\it Confined} \\
    Coupling constant & Strong  & Weak \\ 
    Order parameters & Chiral condensate $\langle \bar q q \rangle$ 
	& Diquark condensate $\langle qq \rangle$\\ 
    Symmetry breaking patterns & $\SU(3)_L \times \SU(3)_R \times \U(1)_B$ 
	& $\SU(3)_C \times \SU(3) _L \times \SU(3)_R \times \U(1)_B$\\
	  & $\rightarrow \SU(3)_{L+R}$ & $\rightarrow \SU(3)_{C+L+R}$\\ 
    Fermions & Octet baryons & Octet + singlet quarks  \\ 
    Vectors & Octet + singlet vector mesons  & Octet gluons  \\
    Nambu-Goldstone modes & Octet pions ($\bar q q$)  
	& Octet + singlet pions ($\bar q \bar q qq $)  \\
	& $H$ boson & $H$ boson \\ \hline
  \end{tabular}
  \end{center}
  \caption{
    Comparisons of the physics between the hadron phase 
    and the CFL phase in massless three-flavor QCD: symmetry breaking patterns 
	[the $\U(1)_A$ and discrete symmetries are suppressed here]
	and the elementary excitations. We have shown
    that confined monopoles appear in the form of the mesonic bound state in the CFL phase.
    Still one missing piece in the table is the properties
    of monopoles in the hadron phase, for which we speculate the
    condensation of monopoles corresponding to the condensation 
    of quarks in the CFL phase.
    See the text for further explanations.}
  \label{tab:continuity}
  \end{table*}

In this subsection, we would like to ask the implications of our results.
In Sec.~\ref{sec:solve}, we found the color-octet mesonic bound states 
formed by of monopole-antimonopole pairs.
Because of the color-flavor locking, they also form flavor-octet under 
the remaining $\SU(3)_{C+F}$.
Clearly, these bound states resemble the flavor-octet mesons formed by
quark-antiquark pairs in the hadron phase.
This leads us to speculate on the idea of the ``quark-monopole duality": 
the roles played by quarks and monopoles are interchanged between
the hadron phase and the CFL phase.
If this is indeed the case, this would imply the condensation 
of monopoles in the hadron phase corresponding to the condensation 
of quarks in the CFL phase. This naturally embodies the
dual superconducting scenario for the quark confinement 
in the hadron phase \cite{N74}.

The possible quark-monopole duality may have some relevance to 
the one-to-one correspondence of the physics without any phase transition
between the hadron phase and the CFL phase conjectured by Sch\"afer and Wilczek \cite{SW99}.
This is called the ``hadron-quark continuity" and may be realized 
in the QCD phase structure in three-flavor limit as explicitly shown 
in \cite{HTYB06, YTHB07}.\footnote{Here we mean the ``hadron phase" 
by the three-flavor symmetric nuclear matter (hyper nuclear matter) where the
$\U(1)_B$ symmetry is dynamically broken by the baryon-baryon pairing.}
The correspondence in the quark-monopole duality and the hadron-quark continuity 
is summarized in Table~\ref{tab:continuity}.
The idea of the hadron-quark continuity is supported by a number 
of nontrivial evidences: the same symmetry breaking patterns, 
the fact that confinement phase is indistinguishable from the Higgs phase \cite{FS79},
the one-to-one correspondence of the elementary excitations such as the baryons, 
vector mesons \cite{HTY08}, and pions \cite{YTHB07},\footnote{The difference 
of the singlet can be ascribed to the mass splitting between the octet and singlet. 
For the quarks in the CFL phase, the singlet is twice heavier than 
the octet [see Eq.~(\ref{eq:spectra})], which is expected to 
correspond to the excited singlet baryonic state 
in the hadron phase. For the vector mesons in the hadron phase, 
the mass splitting is induced by the diquark condensate, and the flavor 
singlet vector disappears at some intermediate $\mu$ \cite{HTY08}.
For the NG modes, the singlet $\eta'$ meson in the hadron phase 
is heavy due to the $\U(1)_A$ anomaly, but becomes a light NG mode 
in the CFL phase by the instanton suppression at large $\mu$ \cite{S01}.
Therefore, the hadron-quark continuity still works.}
and the equivalence of the form of the partition functions
in a finite volume called the $\epsilon$-regime \cite{YK09},
between the hadron phase and the CFL phase.

\begin{center}
\begin{figure}[h]
\includegraphics[width=0.3 \textwidth] {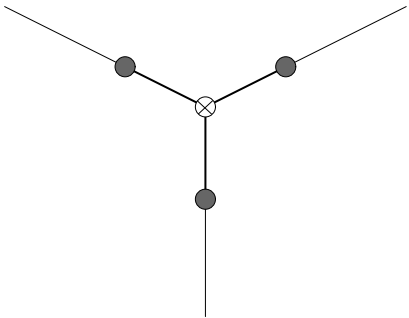}
\caption{Baryonic bound state of three monopoles. 
Three monopoles are connected by a Y-shaped junction of 
color magnetic flux tubes with the total color canceled out.
$\U(1)_B$ vortex is represented by $\otimes$ at the junction point.
This state is dual to a baryon made of three quarks connected 
by a Y-shaped junction of three color electric flux tubes 
in the hadron (confining) phase.}
\label{fig:baryon}
\end{figure}
\end{center}

The quark-monopole duality raises a question regarding possible other 
states formed by monopoles.
In the hadron (confining) phase, a baryonic bound state is made of 
three quarks.
It has been found in \cite{Bali:2000gf} by the lattice QCD simulations that 
three quarks are connected by a Y-shaped junction of color electric flux tubes.
What is the counterpart in the CFL phase?
We expect that it is a junction of three non-Abelian vortices 
with total color fluxes canceled out at the junction point: 
red, blue, and green color magnetic flux tubes 
meet at one point, see Fig.~\ref{fig:baryon}.
We note that they carry correct the baryon number 
as we expect for a baryon;
each flux tube carries the $\U(1)_B$ winding number 1/3, 
and all of them join together to constitute 
one $\U(1)_B$ vortex with the $\U(1)_B$ winding number one.
However we have not specified the electromagnetic charges 
of fluxes at this stage because we have ignored 
the electromagnetic coupling of vortices.
A similar string junction (without monopoles) 
is known to exist in a $\U(1) \times \U(1)$ model~\cite{Bevis:2008hg}.
The configuration in Fig.~\ref{fig:baryon} cannot be discussed 
in the effective field theory 
of a single vortex anymore, but one may be able to do that 
by considering multivortex effective theory.\footnote{
Multivortex states were studied in the 
SUSY QCD \cite{Eto:2005yh}, 
in which case no static interactions exist between 
vortices when they are placed parallel to each other. 
In our case of the CFL phase, parallel vortices are repulsive 
at least when they are well separated \cite{NNM08}.
However short range interactions have not been studied yet, 
and there is a possibility of attraction at short distance.
In any case, we consider that the bound state should be  
quantum mechanically (but not necessary classically) 
stable with the appearance of monopoles, 
as a meson of monopoles found in this paper; 
three vortices with different color fluxes 
join to one $\U(1)_B$ vortex with no fluxes.}

Finally, let us note the effect of the strange quark mass $m_s$
with the charge neutrality and the $\beta$-equilibrium conditions, 
as is expected in the physical dense matter like inside the neutron stars. 
This situation is considered previously without 
the quantum effects for the orientational modes and the nonexistence
of monopoles in the CFL phase is shown \cite{ENY10}.
Even if we further take into account the quantum effects, they are
negligibly-small: the scale of the potential $m_s^2/g_s$ 
for the orientational modes induced by $m_s$ \cite{ENY10},  
is much larger than that induced by the quantum effects $M^2$,
for realistic values of the parameters, $m_s \sim 100 \ {\rm MeV}$, 
$\mu \sim 500 \ {\rm MeV}$, and $\Delta \sim 50 \ {\rm MeV}$;
confined monopoles will be washed out by $m_s$. 
Therefore, we expect that the notion of the quark-monopole duality 
is well-defined close to three-flavor limit.
It is also a dynamical question whether the hadron-quark continuity survives 
when one turns on $m_s$; there are other candidates for the ground state
at intermediate $\mu$ other than the CFL phase under the stress of $m_s$, 
such as the meson condensed phase, 
the crystalline Fulde-Ferrell-Larkin-Ovchinikov phase, 
gluon condensed phase, etc \cite{CSC}.

\section{Conclusion and outlook}
\label{sec:discussion}
In this paper, we have analytically shown that mesonic bound states of 
confined monopoles appear in the color-flavor locked phase 
of three-flavor QCD at large quark chemical potential $\mu$.
They are dynamically generated as kinks by the quantum fluctuations in 
the effective world-sheet theory for the orientational modes on a non-Abelian vortex.
The mass of monopoles has been computed as $M \sim \Delta \exp[- c (\mu/\Delta)^2]$
with the superconducting gap $\Delta$ and some constant $c$.

\begin{center}
\begin{figure}[h]
\includegraphics[width=0.3 \textwidth] {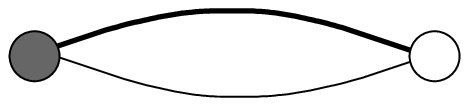} 
\\ (a) \\ \vspace{1cm}
\includegraphics[width=0.3 \textwidth] {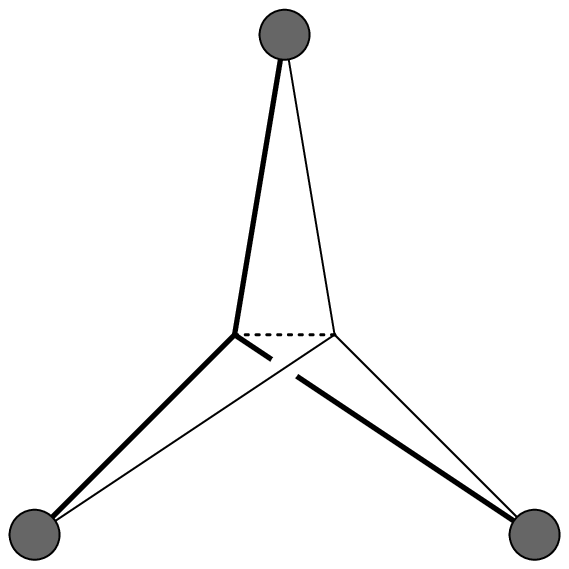} 
\\ (b) 
\caption{
Illustrations of (a) a meson and (b) a baryon 
with finite energy. Flux tubes make loops. 
(a) corresponds to a vortex loop with monopole-antimonopole,
which looks like a necklace.
In (b) three color magnetic fluxes joint at two ends of   
a single $\U(1)_B$ vortex denoted by a broken line. 
}
\label{fig:finite}
\end{figure}
\end{center}

Both of the mesonic and baryonic bound states of monopoles
studied in this paper have long fluxes extending to spatial infinity. 
In a realistic situation these may not be appropriate 
because they have infinite energy. In order for them to have finite energy, 
such long fluxes can be made as loops, 
as illustrated in Fig.~\ref{fig:finite}.\footnote{A 
similar configuration to Fig.~\ref{fig:finite}(a) is also discussed
in the QCD vacuum \cite{C08}.}
For a meson, one can check if this can occur 
by studying the ${\mathbb C}P^2$ model on compactified space $S^1$. 
For a baryon the situation would be more difficult.
A possible configuration for a baryon is given in Fig.~\ref{fig:finite}(b).

Before closing the paper, let us address
several important questions to be investigated in the future.
\begin{enumerate}
\item{
Although we have shown that confined monopoles dynamically appear as 
the monopole-antimonopole mesons by the quantum fluctuations, 
their topological properties are still unclear.
First of all, one should clarify the homotopy group responsible 
for the existence or the topological stability of monopoles.
One should also calculate the color magnetic charge of the confined
monopoles and check if the Dirac condition for the color magnetic 
charge is indeed satisfied. 
The color magnetic flux should be related to the dynamically induced 
$\U(1)_D$ gauge field.
These are in contrast to the situation in the SUSY QCD: 
flux matching between a monopole and a vortex attached to it 
has been demonstrated \cite{Eto:2006dx}.}

\item{Our derivation is based on the effective world-sheet theory on a non-Abelian vortex
derived from the time-dependent Ginzburg-Landau Lagrangian. This is only valid 
near the critical temperature of the CFL phase. One should argue 
the existence of monopoles far away from the critical temperature, e.g., at $T=0$.
As this concerns, the analysis beyond the Ginzburg-Landau theory 
can be studied by the Bogoliubov-de Genne equations 
which describe condensates and quasiparticles 
from the fermion degrees of freedom.
In fact fermion modes have been studied in the presence of 
a non-Abelian vortex by the Bogoliubov-de Genne equations \cite{Yasui:2010yw}.}
\item{
In the case of the SUSY QCD, quantum effects in the (1+1)D on a vortex
can be explained by instanton effects in the original (3+1)D theory
\cite{Shifman:2004dr,SY07}.
Actually, instantons can stably exist inside the vortex world-sheet \cite{Eto:2004rz}.
In real QCD at asymptotic large $\mu$, bulk instanton effects
with the energy $\sim 1/g_s^2 \gg 1$ are highly suppressed due to the
asymptotic freedom of QCD and the screening of instantons \cite{S01}.
The instanton energy in the vortex world-sheet is 
$C_{0,3} \sim (\mu/\Delta)^2 \gg 1/g_s^2$ [see Eq.~(\ref{eq:Kahrer})],
which is further suppressed, 
consistent with our result.
As in the SUSY QCD, the quantum effects inside the vortex
may be explained by instantons trapped in it, which remains as a future problem.
On the other hand, the fact that the instanton energy 
$C_{0,3}\sim (\mu/\Delta)^2$
inside the vortex is larger than the one $\sim 1/g_s^2$ 
in the bulk  implies that instantons are repulsive from vortices.}
\end{enumerate}

\emph{Note added.}---While this work was being completed, we learned that 
A.~Gorsky, M.~Shifman, and A.~Yung \cite{GSY11} have independently found 
very closely related results.

\section*{Acknowledgements}
M.E. is supported by the Special Postdoctoral Researchers Program at RIKEN. 
M.N. is supported in part by Grant-in-Aid for Scientific Research 
(No. 20740141) from the Ministry of Education, Culture, Sports, Science 
and Technology-Japan.
N.Y. is supported by JSPS Postdoctoral Program for Research Abroad. 
 

\end{document}